\documentstyle[preprint,aps]{revtex}
\begin{document}
\title{An Octonionic Gauge Theory}
\draft
\preprint{UM-P-95/09; RCHEP-95/05}
\author{C. C. Lassig\footnote{ccl@physics.unimelb.edu.au}
and G. C. Joshi\footnote{joshi@physics.unimelb.edu.au}}
\address{Research Centre for High Energy Physics, \\
School of Physics, The University of Melbourne, \\
Parkville, Victoria 3052. Australia}
\date{\today}
\maketitle
\begin{abstract}
The nonassociativity of the octonion algebra necessitates a
bimodule representation, in which each element is represented by a left and
a right multiplier. This representation can then be used to generate gauge
transformations for the purpose of constructing a field theory symmetric
under a gauged octonion algebra, the nonassociativity of which appears as
a failure of the representation to close, and hence produces new interactions
in the gauge field kinetic term of the symmetric Lagrangian.
\end{abstract}
\pacs{02.10.Vr, 11.15.-q}

\section{Introduction}
The successful application of real and complex numbers and quaternions in
physics naturally suggests going one step further, and using the
octonions, and indeed many attempts at this have been made (for a history
of hypercomplex numbers in physics see \cite{Gursey}).
Their algebraic structure
and relations to other algebras, most notably their connections to the
exceptional groups, have often been investigated, but the most striking
feature of the octonion algebra is its nonassociativity.

Although this nonassociativity is normally considered incompatible with
physical application, nonassociative structures do arise and have been
handled in the past.
The formulation of quantum mechanics using the Jacobian algebra is inherently
nonassociative, and octonionic algebras have been used in
this application \cite{Jacob}.
Another example is the three-cocycle \cite{Jackiw}, a direct
manifestation of nonassociativity which appears in relation to magnetic
monopoles and may, like one- and two-cocycles, be connected to anomalies
in field theory.
There does not appear to be any compelling reason for nature
to choose a purely associative mathematics for its interactions.

It is worthwhile then not to simply reject octonions on account of their
nonassociativity but rather to investigate the physical implications of
such a feature. In this letter we attempt to construct a gauge theory
based on the octonion algebra in the familiar manner of Yang-Mills theory,
as in \cite{Waldron}.
As is usual, the generators are represented by matrices which necessarily
associate, but here the underlying nonassociativity of the algebra
manifests itself as a failure of the generator algebra to close, a fact
that affects the theory and its phenomenology.

The generators used come from a bimodular representation of the octonions
\cite{Lohmus},
the theory of which is treated generally
in Section II and then specifically applied to the
octonions in Section III. We also look at the freedom of choice of the
octonionic multiplication table
in Section IV, and what effect this may have on the theory.
In Section V we develop our gauge theory, and briefly investigate the
resulting phenomenological effects.

\section{Alternative Algebras and Bimodular Representations}
To begin with, it is necessary to define an {\em associator},
\cite{Lohmus}, \cite{Waldron}, analogous to
the usual algebraic commutator. The associator is
\begin{equation}
\{x, y, z\} = (xy)z - x(yz).
\end{equation}
An {\em alternative algebra} $A$ is defined as an algebra
whose elements satisfy
the property $x^2 y~=~x (x y), (y x) x~=~y x^2, \forall x,y \in A$, or
equivalently, $\{x,x,y\} = \{y,x,x\} = 0, \forall x,y \in A$.
This gives rise to the relation
\begin{equation}
\{x,y,z\} + \{x,z,y\} = 0,
\label{Alt}
\end{equation}
and similarly for other interchanges of $x$, $y$ and $z$.

Because of the nonassociativity of these algebras, it is necessary to use a
{\em bimodular representation}, in which each element $x$ of $A$ is
represented by two linear transformations on $A$, $L_x$ and $R_x$, the left
and right multipliers respectively. These transformations act as
\begin{eqnarray}
& L_x a = & x a, \nonumber\\
& R_x a = & a x, \; \; {\rm where} \; x,y,a \in A.
\end{eqnarray}

{}From the alternativity condition (\ref{Alt}) can then be obtained the
following relations:
\begin{eqnarray}
& & L_{xy} = L_x L_y + [L_x , R_y] \nonumber\\
& & R_{yx} = R_x R_y + [R_x , L_y] \nonumber\\
& & [L_x , R_y] = [R_x , L_y], \; \; \; \; \forall x,y \in A.
\label{Bimod}
\end{eqnarray}

Related to this bimodular representation of multiplications, there are also
two division tables: the {\em left quotient}, $x = b \backslash a$, with $x,a,b
\in A$, such that $b x = a$; and the {\em right quotient}, $y = a/b$ such that
$yb = a$, with $y,a,b \in A$. As we shall see later on, the left and right
representations for the octonions can be obtained from the appropriate
division tables.

\section{The Regular Representation of the Octonion Algebra}

We will now apply the bimodular representation theory defined above to a
particular alternative algebra, that of the {\em octonions}, the last in
the sequence of division algebras of the Hurwitz theorem: real
numbers, complex numbers, quaternions and octonions
\cite{Lohmus}. A general octonion
can be written
\begin{eqnarray}
x & = & x_0 e_0 + x_1 e_1 + x_2 e_2 + x_3 e_3 + x_4 e_4 + x_5 e_5 + x_6 e_6
+ x_7 e_7 \nonumber\\
& = & x_0 e_0 + x_i e_i = x_{\mu} e_{\mu}
\end{eqnarray}
where the $e_{\mu}$ are the octonionic units.
(Here, and elsewhere, we follow the convention that greek indices include the
identity element of the algebra so that $\alpha, \beta, \gamma, \ldots
= 0, \ldots, 7$; latin indices denote only the hypercomplex units so that
$i,j,k, \ldots = 1,\ldots,7$.)
These units satisfy the following multiplication relations:
\begin{eqnarray}
& & e_0^2 = e_0 \nonumber\\
& & e_0 e_i = e_i e_0 = e_i \nonumber\\
& & e_i e_j = -\delta_{ij} e_0 + \epsilon_{ijk} e_k,
\label{Units}
\end{eqnarray}
where $\epsilon_{ijk}$ is totally antisymmetric, and $\epsilon_{ijk} = 1$
for ($ijk$) = (123), (145), (176), (246), (257), (347) and (365) (each cycle
represents a quaternion subalgebra). It also satisfies the relations
\begin{eqnarray}
& & \epsilon_{abi} \epsilon_{cdi} = \delta_{ac} \delta_{bd} - \delta_{ad}
\delta_{bc} + \epsilon_{abcd} \nonumber\\
& & \epsilon_{aij} \epsilon_{bij} = 6 \delta_{ab} \nonumber\\
& & \epsilon_{ijk} \epsilon_{ijk} = 42
\label{Trace}
\end{eqnarray}
where $\epsilon_{abcd}$ is the totally antisymmetric dual tensor, with
$\epsilon_{abcd} = 1$ for ($abcd$) = (1247), (1265), (2345), (2376), (3146),
(3157) and (4576). This tensor also appears in the associator for the units,
\begin{equation}
\{e_i, e_j, e_k\} = 2 \epsilon_{ijkl} e_l.
\end{equation}

The {\em conjugate} $\overline{x}$ of an octonion $x$ may be defined as
$\overline{x} = x_0 e_0 - x_i e_i$.
This mapping $x \rightarrow \overline{x}$ is an {\em involution}, such that
$\overline{\overline{x}} = x$ and
$\overline{xy} = \overline{y} \, \overline{x}$.
The {\em norm} can then be defined,
$|x|^2 = x \overline{x} = \overline{x} x = ( x_0^2 + x_i x_i) e_0$,
as well as a unique inverse,
\begin{equation}
x^{-1} = \frac{\overline{x}}{|x|^2}.
\end{equation}
In terms of the inverse, the left and right quotients can then be written
\begin{eqnarray}
& x = & b \backslash a = b^{-1} a, \\
& y = & a / b = a b^{-1}.
\end{eqnarray}

The multiplication relations Eq.~(\ref{Units}) can also be written in terms
of structure constants $C_{\mu \nu}^{\lambda}$,
\begin{equation}
e_{\mu} e_{\nu} = C_{\mu \nu}^{\lambda} e_{\lambda},
\end{equation}
with $C_{00}^0 = 1$, $C_{i0}^0 = C_{0j}^0 = C_{00}^k = 0$, $C_{ij}^0 =
- \delta_{ij}$, $C_{i0}^k = \delta_{ik}$, $C_{0j}^k = \delta_{jk}$, and
$C_{ij}^k = \epsilon_{ijk}$.
The alternativity condition $\{e_{\mu}, e_{\nu}, e_{\lambda}\} +
\{ e_{\mu}, e_{\lambda}, e_{\nu} \} = 0$ then becomes
\begin{equation}
C_{\nu \lambda}^{\rho} C_{\mu \rho}^{\sigma} e_{\sigma} +
[C_{\lambda \nu}^{\rho} C_{\mu \rho}^{\sigma} e_{\sigma} -
C_{\mu \lambda}^{\rho} C_{\rho \nu}^{\sigma} e_{\sigma}]
= C_{\mu \nu}^{\rho} C_{\rho \lambda}^{\sigma} e_{\sigma}.
\end{equation}
Defining $(C_{\mu})^{\sigma}_{\rho} \equiv C_{\mu \rho}^{\sigma}$, and
$(\tilde{C}_{\nu})^{\rho}_{\lambda} \equiv \tilde{C}_{\nu \lambda}^{\rho}
\equiv C_{\lambda \nu}^{\rho}$, this can be written
\begin{equation}
(C_{\mu})^{\sigma}_{\rho} (C_{\nu})^{\rho}_{\lambda} + \Big[
(C_{\mu})^{\sigma}_{\rho} (\tilde{C}_{\nu})^{\rho}_{\lambda} -
(\tilde{C}_{\nu})^{\sigma}_{\rho} (C_{\mu})^{\rho}_{\lambda} \Big]
= C_{\mu \nu}^{\rho} (C_{\rho})^{\sigma}_{\lambda}.
\label{Reg1}
\end{equation}
Similarly, other relations can be obtained from the remaining
associator conditions.
{}From $\{ e_{\nu}, e_{\lambda}, e_{\mu} \} + \{ e_{\lambda}, e_{\nu}, e_{\mu}
\}
= 0$, and $\{ e_{\nu}, e_{\lambda}, e_{\mu} \} + \{ e_{\mu}, e_{\lambda},
e_{\nu} \} = 0$, we obtain respectively
\begin{eqnarray}
& & \Big[ (\tilde{C}_{\mu})^{\sigma}_{\rho} (C_{\nu})^{\rho}_{\lambda} -
(C_{\nu})^{\sigma}_{\rho} (\tilde{C}_{\mu})^{\rho}_{\lambda} \Big]
+ (\tilde{C}_{\mu})^{\sigma}_{\rho} (\tilde{C}_{\nu})^{\rho}_{\lambda} =
\tilde{C}_{\mu \nu}^{\rho} (\tilde{C}_{\rho})^{\sigma}_{\lambda} \nonumber\\
\mbox{and} \; \; & & \Big[ (\tilde{C}_{\mu})^{\sigma}_{\rho}
( C_{\nu})^{\rho}_{\lambda} - (C_{\nu})^{\sigma}_{\rho}
(\tilde{C}_{\mu})^{\rho}_{\lambda} \Big] +
\Big[ (\tilde{C}_{\nu})^{\sigma}_{\rho} (C_{\mu})^{\rho}_{\lambda} -
(C_{\mu})^{\sigma}_{\rho} (\tilde{C}_{\nu})^{\rho}_{\lambda} \Big] = 0 .
\label{Reg2}
\end{eqnarray}

Using these relations we are able to define a bimodule, the {\em regular
representation}. The left representation is defined $(L_{\mu})_{\lambda \nu}
\equiv (C_{\mu})^{\lambda}_{\nu}$, and the right representation is
$(R_{\mu})_{\lambda \nu} \equiv (\tilde{C}_{\mu})^{\lambda}_{\nu}$. These
matrices then obey the following relations:
\begin{eqnarray}
& & L_{\mu} L_{\nu} + [L_{\mu},R_{\nu}] = C_{\mu \nu}^{\lambda} L_{\lambda},
\nonumber\\
& & R_{\mu} R_{\nu} + [R_{\mu},L_{\nu}] = \tilde{C}_{\mu \nu}^{\lambda}
R_{\lambda}, \nonumber\\
& & [L_{\mu},R_{\nu}] = [R_{\mu},L_{\nu}], \nonumber\\
& & L_{\mu} L_{\nu} + L_{\nu} L_{\mu} = (C_{\mu \nu}^{\lambda} +
C_{\nu \mu}^{\lambda}) L_{\lambda}, \nonumber\\
& & R_{\mu} R_{\nu} + R_{\nu} R_{\mu} = (C_{\mu \nu}^{\lambda} +
C_{\nu \mu}^{\lambda}) R_{\lambda}, \nonumber\\
& & L_{\mu} L_{\nu} + R_{\nu} R_{\mu} = C_{\mu \nu}^{\lambda}
(L_{\lambda} + R_{\lambda}).
\end{eqnarray}
The first three of these relations are simply those above, Eqs~(\ref{Reg1})
and (\ref{Reg2}) rewritten in terms of the left and right matrices, and
correspond to the relations for a general bimodule, Eq.~(\ref{Bimod}).

It is possible to obtain the left and right representation matrices from the
division tables. Writing the elements of the right division table as
$e_{\nu} / e_{\sigma} = K_{\nu \sigma}^{\mu} e_{\mu}$, then from the definition
of the right quotient $K_{\nu \sigma}^{\mu} e_{\mu} e_{\sigma} =
K_{\nu \sigma}^{\mu} C_{\mu \sigma}^{\lambda} e_{\lambda} = e_{\nu}$ (where
the index $\sigma$ is not summed). Since $C_{\mu \sigma}^{\lambda}$ is
non-zero for only one $\lambda$ for every pair $(\mu \sigma)$, and as the units
are linearly independent, $K_{\nu \sigma}^{\mu} = C_{\mu \sigma}^{\nu}$.
Hence $e_{\nu} / e_{\sigma} = C_{\mu \sigma}^{\nu} e_{\mu}$.
In a similar manner, the left division table gives
$e_{\sigma} \backslash e_{\nu} = \tilde{C}_{\mu \sigma}^{\nu} e_{\mu}$.

It is now possible for us to write down these left and right representation
matrices. We shall write them in terms of the Pauli matrices $\sigma_1$,
$\sigma_2$ and $\sigma_3$, with the inclusion of a fourth, $\sigma_0$, merely
the $2 \times 2$ identity matrix:
\begin{equation}
\sigma_0 = \left( \begin{array}{cc} 1 & 0 \\ 0 & 1 \end{array} \right),
\sigma_1 = \left( \begin{array}{cc} 0 & 1 \\ 1 & 0 \end{array} \right),
i\sigma_2 = \left( \begin{array}{cc} 0 & 1 \\ -1 & 0 \end{array} \right),
\sigma_3 = \left( \begin{array}{cc} 1 & 0 \\ 0 & -1 \end{array} \right).
\end{equation}

The left and right matrices are then
\begin{eqnarray}
& L_0 = \left( \begin{array}{cccc} \sigma_0 & 0 & 0 & 0 \\
				   0 & \sigma_0 & 0 & 0 \\
				   0 & 0 & \sigma_0 & 0 \\
				   0 & 0 & 0 & \sigma_0
		\end{array} \right),
& R_0 = \left( \begin{array}{cccc} \sigma_0 & 0 & 0 & 0 \\
				   0 & \sigma_0 & 0 & 0 \\
			           0 & 0 & \sigma_0 & 0 \\
			           0 & 0 & 0 & \sigma_0
       	        \end{array} \right), \nonumber\\
& L_1 = \left( \begin{array}{cccc} -i\sigma_2 & 0 & 0 & 0 \\
				   0 & -i\sigma_2 & 0 & 0 \\
				   0 & 0 & -i\sigma_2 & 0 \\
				   0 & 0 & 0 & i\sigma_2
		\end{array} \right),
& R_1 = \left( \begin{array}{cccc} -i\sigma_2 & 0 & 0 & 0 \\
				   0 & i\sigma_2 & 0 & 0 \\
				   0 & 0 & i\sigma_2 & 0 \\
				   0 & 0 & 0 & -i\sigma_2
		\end{array} \right), \nonumber\\
& L_2 = \left( \begin{array}{cccc} 0 & -\sigma_3 & 0 & 0 \\
				   \sigma_3 & 0 & 0 & 0 \\
				   0 & 0 & 0 & -\sigma_0 \\
				   0 & 0 & \sigma_0 & 0
		\end{array} \right),
& R_2 = \left( \begin{array}{cccc} 0 & -\sigma_0 & 0 & 0 \\
				   \sigma_0 & 0 & 0 & 0 \\
				   0 & 0 & 0 & \sigma_0 \\
				   0 & 0 & -\sigma_0 & 0
		\end{array} \right), \nonumber\\
& L_3 = \left( \begin{array}{cccc} 0 & -\sigma_1 & 0 & 0 \\
				   \sigma_1 & 0 & 0 & 0 \\
				   0 & 0 & 0 & -i\sigma_2 \\
				   0 & 0 & -i\sigma_2 & 0
		\end{array} \right),
& R_3 = \left( \begin{array}{cccc} 0 & -i\sigma_2 & 0 & 0 \\
				   -i\sigma_2 & 0 & 0 & 0 \\
				   0 & 0 & 0 & i\sigma_2 \\
				   0 & 0 & i\sigma_2 & 0
		\end{array} \right), \nonumber\\
& L_4 = \left( \begin{array}{cccc} 0 & 0 & -\sigma_3 & 0 \\
				   0 & 0 & 0 & \sigma_0 \\
				   \sigma_3 & 0 & 0 & 0 \\
				   0 & -\sigma_0 & 0 & 0
		\end{array} \right),
& R_4 = \left( \begin{array}{cccc} 0 & 0 & -\sigma_0 & 0 \\
				   0 & 0 & 0 & -\sigma_0 \\
				   \sigma_0 & 0 & 0 & 0 \\
				   0 & \sigma_0 & 0 & 0
		\end{array} \right), \nonumber\\
& L_5 = \left( \begin{array}{cccc} 0 & 0 & -\sigma_1 & 0 \\
				   0 & 0 & 0 & i\sigma_2 \\
				   \sigma_1 & 0 & 0 & 0 \\
				   0 & i\sigma_2 & 0 & 0
		\end{array} \right),
& R_5 = \left( \begin{array}{cccc} 0 & 0 & -i\sigma_2 & 0 \\
				   0 & 0 & 0 & -i\sigma_2 \\
				   -i\sigma_2 & 0 & 0 & 0 \\
				   0 & -i\sigma_2 & 0 & 0
		\end{array} \right), \nonumber\\
& L_6 = \left( \begin{array}{cccc} 0 & 0 & 0 & -\sigma_0 \\
				   0 & 0 & -\sigma_3 & 0 \\
				   0 & \sigma_3 & 0 & 0 \\
				   \sigma_0 & 0 & 0 & 0
		\end{array} \right),
& R_6 = \left( \begin{array}{cccc} 0 & 0 & 0 & -\sigma_3 \\
				   0 & 0 & \sigma_3 & 0 \\
				   0 & -\sigma_3 & 0 & 0 \\
				   \sigma_3 & 0 & 0 & 0
		\end{array} \right), \nonumber\\
& L_7 = \left( \begin{array}{cccc} 0 & 0 & 0 & -i\sigma_2 \\
				   0 & 0 & -\sigma_1 & 0 \\
				   0 & \sigma_1 & 0 & 0 \\
				   -i\sigma_2 & 0 & 0 & 0
		\end{array} \right),
& R_7 = \left( \begin{array}{cccc} 0 & 0 & 0 & -\sigma_1 \\
				   0 & 0 & \sigma_1 & 0 \\
				   0 & -\sigma_1 & 0 & 0 \\
				   \sigma_1 & 0 & 0 & 0
		\end{array} \right).
\end{eqnarray}

These matrices can be treated as $8 \times 8$ complex matrices, or as
$4 \times 4$ matrices of quaternions, since the Pauli matrices form a
representation of the quaternions.

\section{Modifications of the Multiplication Table}
Due to the arbitrariness in the choice of sign for the antisymmetric tensor
$\epsilon_{ijk}$, it is possible to obtain 16 modifications of the
multiplication table \cite{Lohmus}.
The signs may be arbitrarily chosen for only four
cycles $(ijk)$, then the other three are uniquely determined.
We have chosen to look at the signs of (123), (145), (246) and (347). It
can be shown that
\begin{eqnarray}
\epsilon_{176} & = & \epsilon_{123} \epsilon_{246} \epsilon_{347}, \nonumber\\
\epsilon_{257} & = & \epsilon_{123} \epsilon_{347} \epsilon_{145}, \nonumber\\
\epsilon_{365} & = & \epsilon_{123} \epsilon_{145} \epsilon_{246}.
\end{eqnarray}
Eight of the 16 possible modifications are presented in Table \ref{NoI},
in terms of the relationship of their signs to the
usual multiplication table, known as 0.
As well as these eight there are the {\em inverse} tables, $\tilde{0},
\tilde{1}, \ldots , \tilde{7}$, obtained from the other tables by reversing
the signs of all the cycles.

The numbering system used above comes from the fact that the inverse table
$\tilde{m}$ can be obtained from the default table 0 by the reflection of
an octonionic unit, $e_m \rightarrow -e_m$.
Thus the modified structure constants $C_{\mu \nu}^{(m) \lambda}$ can be
obtained from the unmodified constants by
\begin{eqnarray}
C_{\mu \nu}^{(m) \lambda} & = & C_{\mu \nu}^{\lambda} \; \; \; \; \;
\mbox{if}\; \mu, \nu, \lambda = 0 \; \mbox{or} \; \mu, \nu, \lambda = m,
\nonumber\\
& = & -C_{\mu \nu}^{\lambda} \; \; \mbox{otherwise}.
\end{eqnarray}

The $L_0$ and $R_0$ matrices, being simply the identity matrices, are
unchanged by the modifications. However, the other representation matrices
are affected as follows:
\begin{eqnarray}
L^{(m)}_i & = & \delta_{im} L_m + \epsilon_{imk} L_m L_k, \nonumber\\
R^{(m)}_i & = & \delta_{im} R_m + \epsilon_{imk} R_k R_m,
\end{eqnarray}
where in these equations the index $m$ is not summed.

These different multiplication tables possess the same trace relations
(\ref{Trace}) as before, and hence in the gauge kinetic terms dealt with in
Section V, where a trace is taken over the representation matrices, there
will be no difference. Similarly in coupling to massless fermion or scalar
fields the problem of choice of representation can be overcome by a
redefinition of the fields. It is only when mass terms appear that such a
choice is important, but that is a case of symmetry breaking, and will
not be treated in this letter.

\section{Gauging the Octonion Algebra}
We now wish to construct a field theory invariant under gauge transformations
of the octonion algebra. The nonassociativity of this algebra manifests itself
as a failure of the transformations to close, and so gives rise to new
interactions.

We begin with a matter field
\begin{equation}
\Psi = (\psi^a) = \left( \begin{array}{c} \psi^1 \\ \psi^2 \\ \psi^3 \\ \psi^4
			 \end{array} \right),
\end{equation}
where the entries $\psi^a$ are complex spinors.

To start with, we will consider the gauge transformations
\begin{equation}
\Psi(x) \rightarrow \Psi^{\prime}(x) = e^{i \alpha^i(x) \lambda_i} \Psi(x),
\end{equation}
for which we introduce the matrices $\lambda_i$ and $\rho_j$, defined such
that
\begin{equation}
\lambda_i = \frac{i L_i}4, \; \; \rho_j = \frac{i R_j}4,
\end{equation}
\begin{equation}
[\lambda_i, \lambda_j] = \frac{i}{2} \epsilon_{ijk} \lambda_k
+ 2 [\rho_j, \lambda_i], \nonumber\\
\end{equation}
\begin{equation}
[ \rho_i, \rho_j ] = -\frac{i}{2} \epsilon_{ijk} \rho_k
+ 2 [\lambda_j, \rho_i],
\end{equation}
\begin{equation}
\mbox{Tr}(\lambda_i \lambda_j) = \mbox{Tr}(\rho_i \rho_j) = \frac12
\delta_{ij}.
\label{Trace1}
\end{equation}

Now the usual covariant derivative ${\cal D}_{\mu}$ is introduced,
\begin{equation}
{\cal D}_{\mu} = \partial_{\mu} + i g A_{\mu}(x)
\end{equation}
where $A_{\mu}(x)$ is a member of the octonion algebra and is defined in terms
of gauge fields $A^i_{\mu}(x)$ by $A_{\mu}(x) = A^i_{\mu}(x) \lambda_i$. It
transforms as
\begin{eqnarray}
A_{\mu} \rightarrow A^{\prime}_{\mu} & = & A_{\mu} + [i \alpha^j(x) \lambda_j,
A_{\mu}] - \frac{1}{g} \lambda_i \partial_{\mu} \alpha^i \nonumber\\
& = & A_{\mu} + i \alpha^j A^k_{\mu} \Big( \frac{i}{2} \epsilon_{ijk} \lambda_i
+ 2[\rho_k,\lambda_j] \Big) - \frac1{g} \lambda_i \partial_{\mu} \alpha^i.
\end{eqnarray}
We may split this transformation into closed and non-closed algebraic parts
by denoting $A^{i \prime}_{\mu} \equiv A^i_{\mu} - \frac12 \epsilon_{ijk}
\alpha^j A^k_{\mu} - \frac1g \partial_{\mu} \alpha^i$, so that
\begin{equation}
A_{\mu}^{\prime} = A^{i \prime}_{\mu} \lambda_i + 2 i \alpha^j A^k_{\mu}
[\rho_k, \lambda_j].
\end{equation}

The anti-symmetric curvature tensor is then defined in the usual way as
\begin{eqnarray}
F_{\mu \nu} & = & - \frac{i}{g} [{\cal D}_{\mu}, {\cal D}_{\nu}] \nonumber\\
{} & = & \partial_\mu A_\nu - \partial_\nu A_\mu + i g [ A_\mu , A_\nu ]
\nonumber\\
{} & = & F^i_{\mu \nu} \lambda_i + 2 i g A^j_\mu A^k_\nu [\rho_k, \lambda_j],
\end{eqnarray}
where $F^i_{\mu \nu} \equiv \partial_\mu A^i_\nu - \partial_\nu A^i_\mu
- \frac12 g \epsilon_{ijk} A^j_\mu A^k_\nu$ is the closed part. By using trace
relations Eq.~(\ref{Trace1}) and
\begin{eqnarray}
\mbox{Tr}\Big( \lambda_i [\rho_j, \lambda_k] \Big) & = &
\frac{i}8 \epsilon_{ijk}, \nonumber\\
\mbox{Tr}\Big( [\lambda_i, \rho_j] [\lambda_k, \rho_l] \Big) & = &
- \frac1{32} \Big[ \epsilon_{ijkl} + 2 (\delta_{ik} \delta_{jl} -
\delta_{il} \delta_{jk}) \Big],
\end{eqnarray}
we can write down the gauge field kinetic term as
\begin{eqnarray}
{\cal L}_L & = & - \frac12 \mbox{Tr}(F_{\mu \nu} F^{\mu \nu}) \nonumber\\
{} & = & - \frac14 F^i_{\mu \nu} F^{i \mu \nu} - \frac14 g F^i_{\mu \nu}
A^{j \mu} A^{k \nu} \epsilon_{ijk} - \frac18 g^2 A^j_\mu A^k_\nu
(A^{j \mu} A^{k \nu} - A^{k \mu} A^{j \nu}) \nonumber\\
{} & = & - \frac14 (\partial_\mu A^i_\nu - \partial_\nu A^i_\mu )
(\partial^\mu A^{i \nu} - \partial^\nu A^{i \mu}) - \frac{1}{16} g^2
A^j_\mu A^k_\nu (A^{j \mu} A^{k \nu} - A^{k \mu} A^{j \nu}).
\end{eqnarray}
The Feynman rules obtained from this Lagrangian lack a three--gauge boson
vertex, a fact that could adversely affect the renormalizability of the
theory.

So far we have only gauged the left representation, but it is of course
possible to use the right bimodule instead,
in terms of the transformation $\Psi(x) \rightarrow
\Psi^{\prime}(x) = \exp [i \beta^i(x) \rho_i] \Psi(x)$. We then define the
gauge fields $B^i_\mu (x)$ such that ${\cal D}_\mu = \partial_\mu + igB_\mu(x)$
where $B_\mu (x) = B^i_\mu (x) \rho_i$, with the transformation rule
$B_\mu \rightarrow B^\prime_\mu = \rho_i B^{i \prime}_\mu + 2 i \beta^j B^k_\mu
[\lambda_k, \rho_j]$, where $B^{i \prime}_\mu = B^i_\mu + \frac12
\epsilon_{ijk}
\beta^j B^k_\mu - \frac1g \partial_\mu \beta^i$. The gauge kinetic term is now
\begin{equation}
{\cal L}_R = -\frac14 G^i_{\mu \nu} G^{i \mu \nu} + \frac14 G^i_{\mu \nu}
B^{j \mu} B^{k \nu} \epsilon_{ijk} - \frac18 g^2 B^j_\mu B^k_\nu
(B^{j \mu} B^{k \nu} - B^{k \mu} B^{j \nu}),
\end{equation}
where the field strength tensor $G^i_{\mu \nu} = \partial_\mu B^i_\nu
- \partial_\nu B^i_\mu + \frac12 g \epsilon_{ijk} B^j_\mu B^k_\nu$.

We can now use the above results to generalize to a theory that will be
symmetric with respect to both left and right transformations.
To this end we introduce a new covariant derivative
${\cal D}_\mu = \partial_\mu + ig [A_\mu(x) + B_\mu(x)]$,
with $A_\mu(x)$ and $B_\mu(x)$ defined as previously. The combined
field strength tensor $F_{\mu \nu}$ will now be, in terms of the tensors
$F^i_{\mu \nu}$ and $G^i_{\mu \nu}$ used above,
\begin{equation}
F_{\mu \nu} = F^i_{\mu \nu} \lambda_i + G^i_{\mu \nu} \rho_i
+ ig (2 A^j_\mu A^k_\nu + 2 B^j_\mu B^k_\nu - A^j_\mu B^k_\nu - B^j_\mu A^k_\nu
) [\rho_k, \lambda_j].
\end{equation}
The gauge kinetic term is somewhat more complicated this time,
and is found to be
\begin{eqnarray}
{\cal L}_{LR} & = & -\frac14 F^i_{\mu \nu} F^{i \mu \nu} -
\frac14 G^i_{\mu \nu} G^{i \mu \nu} + \frac14 F^i_{\mu \nu} G^{i \mu \nu} -
\frac14 g F^i_{\mu \nu} (A^{j \mu} A^{k \nu} + B^{j \mu} B^{k \nu} -
A^{j \mu} B^{k \nu} ) \epsilon_{ijk} \nonumber\\
 & & + \frac14 g G^i_{\mu\nu} (A^{j\mu} A^{k\nu} +
B^{j \mu} B^{k \nu} - A^{j\mu} B^{k\nu}) \epsilon_{ijk}
- \frac{3}{32} g^2 A^i_\mu A^j_\nu B^{k\mu} B^{l\nu} \epsilon_{ijkl}
\nonumber\\
& & - \frac18 g^2 A^j_\mu A^k_\nu (A^{j\mu} A^{k\nu} - A^{k\mu} A^{j\nu})
 + \frac1{16} g^2 (4 A^j_\mu A^k_\nu - A^j_\mu B^k_\nu)
(A^{j\mu} B^{k\nu} - A^{k\mu} B^{j\nu})
\nonumber\\
& & - \frac1{16} g^2 (3 A^j_\mu A^k_\nu + 2 B^j_\mu B^k_\nu
- 4 A^j_\mu B^k_\nu)(B^{j\mu} B^{k\nu} - B^{k\mu} B^{j\nu}).
\end{eqnarray}

In order to find the Feynman rules for this theory, it is necessary to write
the Lagrangian in terms of dynamical fields, obtained by diagonalising the
quadratic terms. This is done by putting $A^i_\mu = \frac1{\sqrt{3}} X^i_\mu
+ Y^i_\mu$ and $B^i_\mu = - \frac1{\sqrt{3}} X^i_\mu + Y^i_\mu$, where
$X^i_\mu$ and $Y^i_\mu$ are the new dynamical fields. The Lagrangian then
becomes
\begin{eqnarray}
{\cal L}_{LR} & = & -\frac14 (\partial_\mu X^i_\nu - \partial_\nu X^i_\mu )
(\partial^\mu X^{i \nu} - \partial^\nu X^{i \mu})
- \frac14 (\partial_\mu Y^i_\nu - \partial_\nu Y^i_\mu)
(\partial^\mu Y^{i \nu} - \partial^\nu Y^{i \mu}) \nonumber\\
& & - \frac2{\sqrt3} g \epsilon_{ijk} X^i_\mu X^j_\nu \partial^\mu X^{k \nu}
+ \frac2{\sqrt{3}} g \epsilon_{ijk} (Y^i_\mu Y^j_\nu \partial^\mu X^{k \nu}
+ Y^i_\mu X^j_\nu \partial^\mu Y^{k \nu} + X^i_\mu Y^j_\nu \partial^\mu
Y^{k \nu} ) \nonumber\\
& & - \frac1{16} g^2 (\delta_{ik} \delta_{jl} - \delta_{il} \delta_{jk} )
\Big( \frac79 X^i_\mu X^j_\nu X^{k \mu} X^{l \nu} + \frac23 X^i_\mu X^j_\nu
Y^{k \mu} Y^{l \nu} + 2 X^i_\mu Y^j_\nu X^{k \mu} Y^{l \nu} \nonumber\\
& & + \frac23 X^i_\mu Y^j_\nu Y^{k \mu} X^{l \nu} - Y^i_\mu Y^j_\nu Y^{k \mu}
Y^{l \nu} \Big) + \frac18 g^2 \epsilon_{ijkl}
X^i_\mu X^j_\nu Y^{k \mu} Y^{l \nu}.
\end{eqnarray}
This time there is also some doubt about the renormalizability of the theory,
as for instance there will be no vertex with three $Y$ bosons. To say with
certainty whether this is a problem requires a full analysis
to at least one-loop level,
including the fermion fields as well, and this has yet to be done. Of
course, if supersymmetry is included there will be no problem, but as this
is only a toy model the details of its possible physical applications are
unknown.

\begin{table}
\caption{Signs of $\epsilon_{ijk}$ for four cycles $(ijk)$,
for eight different multiplication tables.}
\label{NoI}
\begin{tabular}
{ccccccccc}
Cycle & 0 & 1 & 2 & 3 & 4 & 5 & 6 & 7 \\
\hline
(123) & + & + & + & + & $-$ & $-$ & $-$ & $-$ \\
(145) & + & + & $-$ & $-$ & + & + & $-$ & $-$ \\
(246) & + & $-$ & + & $-$ & + & $-$ & + & $-$ \\
(347) & + & $-$ & $-$ & + & + & $-$ & $-$ & +
\end{tabular}
\end{table}

\end{document}